\documentclass{aastex63}
\newcommand{\pd}[2]{\displaystyle \frac{\partial #1}{\partial #2}}

\begin{document}



\title{Dynamic evolution of major element chemistry in protoplanetary disks and its implications for chondrite formation}

\correspondingauthor{Yoshinori Miyazaki}
\email{yoshinori.miyazaki@yale.edu}

\author[0000-0001-8325-8549]{Yoshinori Miyazaki}
\affil{Department of Earth and Planetary Sciences, Yale University \\
New Haven, CT 06511 USA}

\author[0000-0002-4785-2273]{Jun Korenaga}
\affil{Department of Earth and Planetary Sciences, Yale University \\
New Haven, CT 06511 USA}


\begin{abstract}
Chondrites are the likely building blocks of Earth, and identifying the group of chondrite that best represents Earth is a key to resolving the state of the early Earth. The origin of chondrites, however, remains controversial partly because of their puzzling major element compositions, some exhibiting depletions in Al, Ca, and Mg.
Based on a new thermochemical evolution model of protoplanetary disks, we show that planetesimals with depletion patterns similar to ordinary and enstatite chondrites can originate at 1--2~AU just outside where enstatite evaporates. Around the ``evaporation front'' of enstatite, the large inward flow of refractory minerals, including forsterite, takes place with a high pebble concentration, and the loss of those minerals result in depletion in Al, Ca, and Mg. When evaporated solid grains re-condense onto pebbles, the concentration of pebbles is further enhanced, potentially triggering the streaming instability. Planetesimals with the composition of ordinary and enstatite chondrites can thus be naturally created in the terrestrial region. The preferential loss of forsterite also creates an enhancement of Mg/Si and heavy Si isotopes just inside the potential source region for ordinary and enstatite chondrites. Earth, which shows both features, may originate just inside where ordinary and enstatite chondrites were born. 
\end{abstract}

\section{Introduction}
Chondrites are a promising candidate for the building blocks of Earth, providing important constraints on the bulk chemistry, volatile inventory, and redox state of Earth's primitive mantle \citep{McDonough1995,Lyubetskaya2007,Dauphas2017}. There has been a long-standing debate over which chondrite group is the best representative of Earth, but no group perfectly matches the mantle in terms of both chemical and isotopic compositions \citep{Jagoutz1979,Allegre1995,Javoy1995}. Among different chondrite groups, enstatite chondrites are isotopically the closest to Earth \citep{Clayton1984a,Dauphas2004}, whereas carbonaceous chondrites are the most similar in terms of the bulk composition \citep{Wasson1988,Halliday2015}. 
One of the difficulties in identifying the building blocks arises from our limited understanding of the mechanisms that are responsible for the major element compositions of chondrites. Ordinary and enstatite chondrites are depleted in refractory elements including Al, Ca, and Mg, but such depletions are enigmatic from the perspective of the volatility trend \citep{Lodders2003,Albarede2014}. The nebular gas dissipates as the protoplanetary disk cools down, so depletion is expected for volatile elements, not for refractory elements \citep{Cassen1996,Ciesla2008}. Also, the enrichment of moderately volatile elements including Na, with respect to Al, Ca, and Mg, is observed in the EH class of enstatite chondrites, making their origin even more puzzling \citep{Wasson1988}.

The diverse compositions of chondrites have been explained by different proportions of their constituents. Chondrites contain four major components: Ca-Al-rich inclusion (CAI), chondrule, metallic Fe, and matrix. The loss of CAIs has been proposed to explain the depletion of refractory elements including Al and Ca. This is corroborated by astrophysical models as CAIs are likely lost to the Sun within the timescale of nebular evolution \citep{Cuzzi2003, Pignatale2018b}. The loss of CAIs, however, is not sufficient to explain the degree of Al- and Ca-depletion in ordinary and enstatite chondrites. The relative amount of Al that is present in CAIs is less than 10\% for carbonaceous chondrites \citep{Hezel2008}, so even losing all of CAIs accounts for less than half of observed depletion. Instead, the other constituents of chondrites including chondrules must have formed in regions where refractory elements were already somehow depleted \citep{Scott2014}.  
The depletion of Mg is even more difficult to explain. The proportions of Mg and Si that are present in CAIs are less than a few percent, and the loss of CAIs cannot account for the observed 10--20\% depletion in Mg. Some attribute the depletion of Mg in enstatite chondrites to some secondary process \citep{Lehner2013}, but such an explanation may not be universally applicable because the proposed secondary process is unlikely to have taken place in ordinary chondrites, which also exhibit similar but weaker Mg-depletion \citep{Wasson1988}.

One possible explanation for the depletion of refractory elements is that Al-, Ca-, and Mg-rich minerals are all preferentially delivered to the inner region of the Solar System, whereas the transport of other minerals is prohibited by evaporation \citep{Miyazaki2017}. Al- and Ca-rich minerals have condensation temperatures of 1600--1800~K at $10^{-3}$~atm, whereas the major component of silicate minerals, enstatite (MgSiO$_3$), condenses below 1400~K \citep{Grossman1972, Lodders2003}. Al- and Ca-rich minerals can thus survive the inward transport within a hot region, which may result in a larger loss of Al and Ca compared to Si. 
On the other hand, fractionation between Mg and Si is likely to be driven by the radial transport of forsterite (Mg$_2$SiO$_4$), which has a larger Mg/Si ratio than enstatite. The preferential loss of forsterite is considered to be consistent with the mineral assemblage of enstatite chondrites \citep{Petaev1998, Grossman2008}. The condensation temperature of forsterite, however, is higher only by $\sim$100~K compared to that of enstatite, and with such a small temperature difference, it is not obvious whether forsterite can decouple from enstatite to create the depletion of Mg observed in ordinary and enstatite chondrites. We need an astrophysical model to investigate how and where the loss of these minerals can happen within protoplanetary disks. 
 
Minerals not only evaporate, but they also react with the surrounding gas and change their compositions while advecting through the protoplanetary disk. When enstatite drifts inward, for instance, it first dissociates to yield forsterite and SiO before its complete evaporation. Solving for the physical and chemical evolution of the disk simultaneously is, therefore, a key to modeling fractionation among major elements. In this study, we incorporate condensation theory into the astrophysical model of protoplanetary disks in a self-consistent manner. Our model tracks actual gaseous and mineral species, whereas previous disk models described the chemical evolution using simplified species, by grouping several minerals into a single component as ``silicates'' or ``volatiles''  \citep{Ciesla2006, Estrada2016, Pignatale2018b}. These simplified components are assumed to evaporate above their condensation temperatures, but such a crude treatment cannot account for fractionation between Mg and Si. Also, by using actual mineral species, we can better predict the Al/Si and Ca/Si ratios because Al- and Ca-rich minerals contain some amount of Mg and Si. Grouping minerals into Al- and Ca-bearing refractories and Mg- and Si-bearing silicates cannot accurately estimate the evolution of Al/Si and Ca/Si ratios.

With our new approach, we show that fractionation between Mg and Si can indeed be driven by the loss of forsterite in the region where enstatite evaporates under certain conditions. 
For the cases where Mg-depletion occurs, the enrichment of dust happens in the same region, potentially triggering the streaming instability \citep{Johansen2007,Johansen2009}, and the resulting planetesimals have the Al, Ca, Mg, and Na/Si ratios close to ordinary and enstatite chondrites. Therefore, our model naturally produces parental bodies for these two classes of chondrites and suggests a new evolutionary path towards forming planetesimals in the terrestrial region.

\section{Model}
Our model solves for the radial motions of all species using the following advection-diffusion equation \citep{Estrada2016, Desch2017}, which includes  an additional term, $S_i$, to account for condensation ($S_i>0$) and evaporation ($S_i<0$): 
\begin{equation} \label{addiff}
	\pd{\Sigma_i}{t} = \frac{1}{r} \pd{}{r} \left( - r v_i \Sigma_i  \right) + \frac{1}{r} \pd{}{r} \left( r \nu \Sigma \pd{}{r} \left( \frac{\Sigma_i}{\Sigma} \right) \right) + S_i.
\end{equation}
The first term on the right hand side describes advection, and the second term is dissipation based on the concentration gradient, where $\Sigma_i$ denotes the surface density of species $i$, $t$ is time, $r$ is the distance from the Sun, $v_i$ is the advection velocity, $\nu$ is viscosity, and $\Sigma$ is the total surface density. 
The species in our model include actual gaseous species and minerals, and the source term $S_i$ is calculated through Gibbs energy minimization \citep[][see Methods for details]{Miyazaki2017}, allowing us to model various kinds of fractionation including the one between Mg and Si. We do not specify the condensation temperature of each species a prioiri. Condensation temperatures are naturally determined through Gibbs energy minimization, and their values vary with pressure and composition.

Gas, dust, and pebble phases are considered for each species, which are used to represent different dynamical behaviors. The pebble phase represents solids with larger grain sizes ($\ge$0.1~mm) than those in the dust phase ($<$0.1~mm). This is motivated by the recent findings that some fraction of solids grow orders of magnitude larger than the main population \citep{Windmark2012,Estrada2016}; larger grains have lower probabilities of destruction and can grow by colliding with smaller grains. The most notable role of such pebbles is their ability to transfer solid components towards the inner region because of large grain sizes and thus large drift velocities \citep{Ciesla2006, Ciesla2008}.
The size and mass fraction of pebbles are therefore important in characterizing fractionation among major elements, but they depend on various factors including turbulent viscosity, specific energies for bouncing and fragmentation, and the efficiency of erosion \citep{Zsom2010, Windmark2012, Estrada2016}. In this study, instead of solving the collisional coagulation equation \citep{Dullemond2005}, we characterize the formation of pebbles using two parameters: the time scale for the dust to grow into pebbles, $\tau_{d \to p}$, and the normalized grain size of pebbles, $\mathrm{St}_p$, to understand the role of pebbles on major element chemistry. For simplicity, we assume a single grain size for the pebble phase, and the consequence of having a distribution in grain size is discussed later. The composition of the pebble phase is assumed to be in equilibrium with the gas and dust phases because the timestep in our model is longer than sublimation time scales for silicates \citep{Takigawa2009}.
We investigate under what conditions protoplanetary disks can create a depletion pattern consistent with ordinary and enstatite chondrites. 

\section{Results}
\subsection{Fractionation between Mg and Si}
\begin{figure}[tbhp]
\centering
\includegraphics[width=13cm]{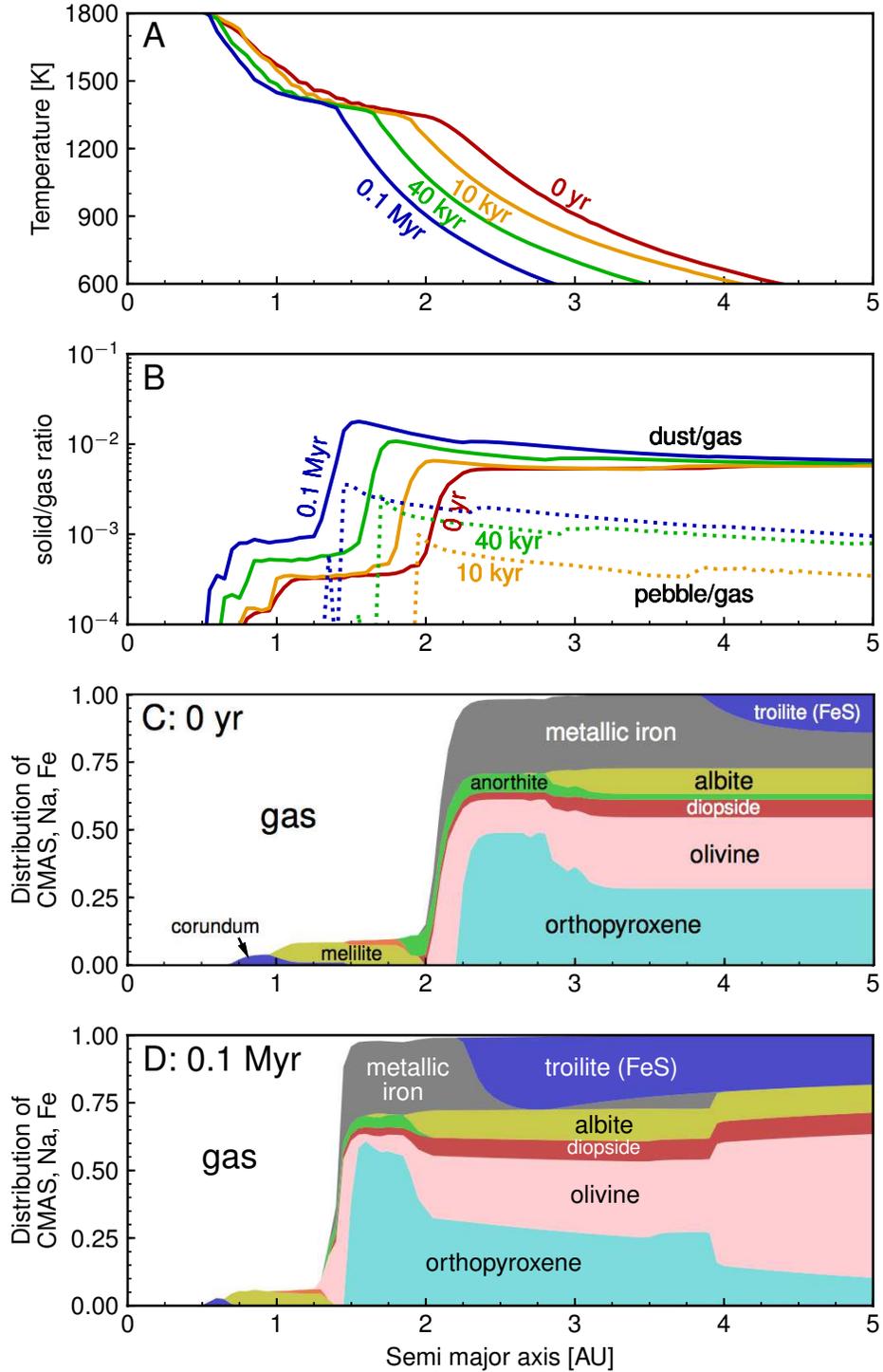}
\caption{(A, B) Snapshots of disk evolution for (A) temperature and (B) the solid/gas ratio. Colors denote 0~kyr (red), 10~kyr (orange), 40~kyr (green), and 0.1~Myr (blue). Solid and dotted lines in (B) indicate the ratios of dust/gas and pebble/gas, respectively. (C, D) Predicted mineral assemblages of the system at (C) $t$=0~yr and (D) $t$=0.1~Myr. The relative molar amount of solid-composing elements, Na, Mg, Al, Si, Ca, and Fe, in the solid phase is plotted as a function of heliocentric distance, where colors denote different mineral phases: corundum (Al$_2$O$_3$, dark blue), melilite (Ca$_2$Al$_2$SiO$_7$, yellow), orthopyroxene ((Mg,Fe)SiO$_3$, cyan), olivine ((Mg,Fe)$_2$SiO$_4$, pink), diopside (CaMgSi$_2$O$_6$, red), anorthite (CaAl$_2$Si$_2$O$_8$, green), albite (NaAlSI$_3$O$_8$, yellow), metallic iron (Fe, gray), and troilite (FeS, purple).}
\label{fig_Tdr}
\end{figure}

Our model is the first to dynamically track the evolution of major element chemistry within protoplanetary disks. The physical evolution of the disk is similar to previous models \citep{Ruden1991, Estrada2016}, where temperature gradually decreases, and solid is enriched in the inner region as pebbles are transported inward by radial drifting (Figure~\ref{fig_Tdr}a, b). Mineral assemblages predicted in our model, however, indicate the new aspects of the protoplanetary disk chemistry. They change mostly as a function of temperature, but in regions where minerals evaporate, constituent elements are enriched by the transport of pebbles, altering mineral compositions. The most notable change is regarding Mg- and Si-bearing minerals in the region where enstatite evaporates (which we hereafter call the evaporation front of enstatite). The ratio between enstatite (orthopyroxene) and forsterite (olivine) decreases from $\sim$0.17 to $\sim$0.05 after 0.1~Myr of evolution (Figure~\ref{fig_Tdr}c, d). The value of 0.05 is consistent with EH chondrites \citep{Nehru1984}, and this study is the first to reproduce this value in the context of astrophysics. Because the molar Mg/Si ratio of forsterite is 2:1 and is larger than that of enstatite, fractionation between forsterite and enstatite results in the depletion of Mg (Figures~\ref{fig_MgSi}A, \ref{fig_MgSi}C). 

Such enrichment of enstatite can be explained by the preferential loss of forsterite, which is driven by the difference in condensation temperatures between forsterite and enstatite. Because enstatite evaporates $\sim$100~K lower than forsterite, the evaporation front of enstatite encloses the region where forsterite remains as solid (Figure~\ref{fig_Tdr}C). While enstatite evaporates, condensed forsterite continues to be incorporated into pebbles and is transported further to the inner region, leaving the region outside the evaporation front to be depleted in Mg. Pebbles thus hold a key to creating fractionation; our model produces negligible fractionation when the pebble/gas ratio is low. A shorter time scale for pebble formation produces a higher surface density of pebbles, resulting in a greater degree of fractionation (Table~\ref{tab_res}). 

A high concentration of pebbles is required because chemical heterogeneity created by pebble transport can be eliminated by vigorous turbulence within the protoplanetary disk. Because enstatite and forsterite evaporate at nearby regions (Figures~\ref{fig_Tdr}C, \ref{fig_Tdr}D), fractionation between Mg and Si is more vulnerable to turbulence and is easier to be re-homogenized than fractionation between other elements. 
We use the $\alpha$-prescription of turbulent viscosity \citep{Shakura1973} to describe the strength of turbulence, and fractionation between Mg and Si large enough to explain ordinary chondrites is produced only for $\alpha \le 10^{-3}$ (Figures~\ref{fig_MgSi}A, \ref{fig_MgSi}C). For the cases of $\alpha= 10^{-2}$, which is the upper estimate for protoplanetary disks \citep[e.g.,][]{Davis2010, Armitage2011}, fractionation between Mg and Si is limited (Figures~\ref{fig_MgSi}E, S3A). The degree of fractionation is too low ($\sim$2\%) compared to what is observed in chondrites \citep[$\sim$10-18\%,][]{Wasson1988}, so the value of $10^{-2}$ would be most applicable to our solar system if the suggested mechanism for the Mg-depletion is incorrect.

The other parameter investigated in our model is the size of pebbles, which is characterized by the normalized grain size St$_p$. The advection velocity of pebbles, being proportional to their grain size for St$_p$$<$1, is directly related to the ability to produce chemical heterogeneity. Therefore, larger pebbles could induce more fractionation between Mg and Si (Figure~\ref{fig_MgSi}C) even with a lower pebble/gas ratio.
When the other parameters are the same, however, a larger pebble size lowers the concentration of pebbles because pebbles drift faster and are lost towards the Sun before their concentration increases. In regions not affected by condensation and evaporation, the pebble/gas ratio is controlled by a balance between the production of pebbles and their loss towards the inner region by radial drift. Therefore, having a large pebble size does not always result in a higher degree of depletion (Al/Si and Na/Si in Figures~\ref{fig_MgSi}A, \ref{fig_MgSi}C). 

\subsection{Depletion of other refractory elements}
Highly refractory elements including Al and Ca show a larger depletion compared to Mg. The condensation temperatures of Al- and Ca-bearing minerals such as corundum and melilite are higher than that of enstatite by 200--300~K, which corresponds to a $\sim$1~AU difference in the distance from the Sun for the case shown in Figure~\ref{fig_Tdr}. Because the length scale of turbulent diffusion is shorter than 1~AU, turbulent mixing does not eliminate fractionation among Al, Ca, and Si, resulting in a larger degree of depletion compared to Mg. 

\begin{figure}[tbhp]
\begin{center}
\includegraphics[width=19cm]{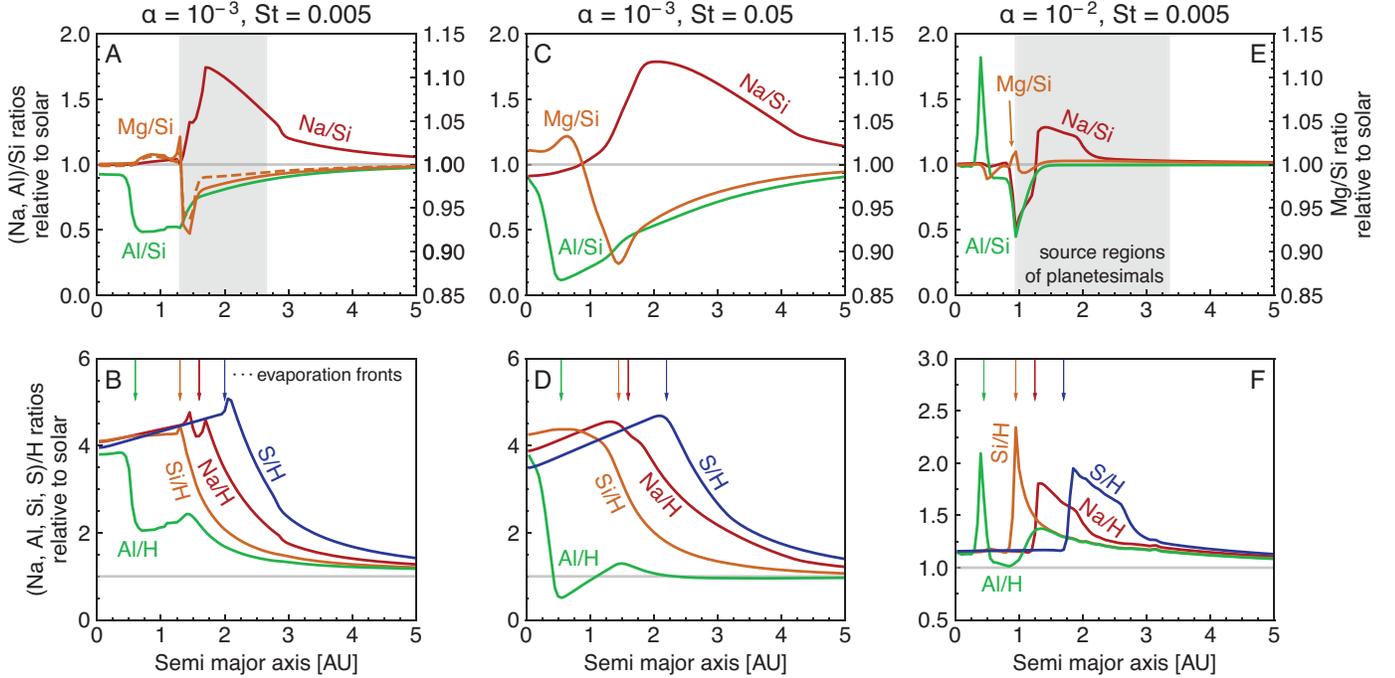}
\end{center}
\caption{Modeled profiles of system composition in terms of the relative abundances of (A, C, E) Na (red), Mg (orange), and Al (green) normalized by Si and the composition of solar composition, and (B, D, F) Na (red), Al (green), Si (orange), and S (blue) normalized by H and the solar composition.  These plots show the profile after 0.1~Myr of evolution for (A, B) and 40~kyr for (C). The parameters used to create this figure is the same with Figure~\ref{fig_Tdr} for (A, B). Cases when (C, D) the normalized pebble size is 10 times larger and when (E, F) the turbulent viscosity is 10 times larger are also shown. The dotted line in (A) indicates the case where the re-condensation of solid-composing elements onto pebbles is neglected. Gray lines in all panels denote unity, corresponding to an unfractionated value. Regions where planetesimals may form by the streaming instability are shaded in (A, E). In (C), the pebble/gas ratio is too low to trigger the streaming instability, so no region is shaded. In (B, D, F), the evaporation fronts of corundum (green), enstatite (orange), albite (red), and troilite (blue) are denoted by arrows. These minerals are the major host phases of Al, Si, Na, and S, respectively, at the time of evaporation.}
\label{fig_MgSi}
\end{figure}

The Al/Si ratio of the system composition, normalized by the solar composition, is lower than unity in the inner region of the disk, but elements that constitute solids are mostly enriched when compared to H (Figures~\ref{fig_MgSi}B, \ref{fig_MgSi}D, and \ref{fig_MgSi}F). The Al/Si ratio can thus be used as the most sensitive proxy for the degree of enrichment. Solid-composing elements are enriched around the evaporation front of each mineral through the continuing transport and evaporation of pebbles. The most refractory element in this study, Al, shows enrichment in the region closest to the Sun, and other peaks of enrichment are observed further away from the Sun for elements with higher volatility (Si, Na, and S in the increasing order of volatility). In the region around the evaporation front of enstatite, Si is enriched but not so much for Al, so the ratio of Al/Si shows a significant decrease. The enrichment of moderately volatile elements can be explained similarly. 
The ratio of Na/Si, for example, is higher than unity outside the evaporation front of enstatite because Na is enriched but not Si. Solids formed from further away from the Sun thus would be depleted in Al and Mg, but enriched in Na compared to Si. If final planetary bodies contain some amount of these Na-rich solids, it may explain the enigmatic volatile enrichment observed in the EH class of enstatite chondrites \citep[][see Discussion]{Wasson1988}.

\subsection{The enhancement of depletion}
An additional mechanism that enhances the degree of depletion is the re-condensation of evaporated solid-composing elements on the pebble phase. When pebbles cross the evaporation front, some of the constituent elements are partitioned into the gas phase. A fraction of them, however, are transported back to the outer region by turbulent diffusion and re-condense as solid. In our model, we have distributed the amount of re-condensation to the dust and pebble phases according to their total surface areas. Previous models have assumed that all of the re-condensed solid turns into the dust of small grain sizes \citep{Ciesla2006, Estrada2016}, but this would underestimate the role of pebbles. In terms of chemistry, the degree of Mg-depletion increases by $\sim$40\% (Mg/Si going down from 0.93 to 0.9; Figure~\ref{fig_MgSi}A), which is comparable to what is observed in ordinary chondrites (Mg/Si$\sim$0.9). The re-condensation onto pebbles also affects the pebble density outside the evaporation front. The region just outside the evaporation front of enstatite is affected most, and the pebble density can increase by a factor of 2--3.

When the ratio between the surface densities of pebbles and gas reaches $\sim$1.5\%, the streaming instability can be triggered, which is a gravitational instability induced by aerodynamic coupling between solids and gas \citep{Johansen2007, Johansen2009}. The concentration of silicate pebbles was never high enough to trigger the streaming instability in previous models \citep{Cuzzi2004,Estrada2016}, but with the re-condensation of dust components, the pebble/gas ratio can become very close to this threshold (Figure~\ref{fig_Tdr}B). This may suggest a new pathway towards creating planetesimals in the inner region of protoplanetary disks. The region just outside the evaporation front of enstatite, where the pebble/gas ratio is sufficiently high, is dominated by enstatite and is depleted in forsterite (Figure~\ref{fig_Tdr}D). Depletion in Al, Ca, and Mg is consistent with ordinary and enstatite chondrites, indicating that planetesimals created by the streaming instability outside the evaporation front of enstatite may serve as a source for these chondrites. 

\newpage
\section{Discussion}
\subsection{Constraints on astrophysical parameters}
Because fractionation between Mg and Si occurs only under limited conditions, our model may be used to constrain astrophysical parameters using cosmochemical observations. To produce Mg-depletion observed in ordinary chondrites, the following three conditions should be met (Table~\ref{tab_res}): turbulent mixing is sufficiently weak ($\alpha \le 10^{-3}$), the time scale of pebble formation is short ($\tau_{d \to p} \le 100$~yr), and the grain size of pebbles is in the range corresponding to St$_p \sim$0.01 (e.g., 10~cm at 2~AU and $t$$\sim$$10^4$~yr). 
Strong turbulence is not preferred by our results, so magnetorotational instability (MRI), for which the suggested value of $\alpha$ is on the order of $10^{-2}$ \citep{Balbus1991}, may not have been the source of turbulence in the Solar System. Because the inner region was sufficiently hot and partially ionized during the earlier stage of evolution, MRI has been considered as the primary source of turbulence, but a mechanism that produces weaker turbulence, including purely hydrodynamical turbulence, is instead suggested from the Mg-depletion observed in ordinary and enstatite chondrites. 

The size of pebbles may also be constrained using the major element chemistry. When the normalized pebble size St$_p$ is smaller than $10^{-3}$, fractionation between Mg and Si becomes too small to explain the composition of chondrites. A larger pebble size, however, is not necessarily favored as well because it decreases the concentration of pebbles, lowering the possibility of the streaming instability \citep{Youdin2005, Johansen2009}. Although greater fractionation is observed when St$_p$ exceeds $10^{-2}$ (Figure~\ref{fig_MgSi}C), the pebble/gas ratio becomes smaller than $10^{-3}$ (Figure~\ref{fig_Tdr_32294}). The sweet spot of the pebble size is likely to be around St$_p \sim$ 0.005, where Mg-Si fractionation and the large pebble concentration are both achieved. In reality, the size of pebbles should exhibit a certain degree of distribution, and the compositional evolution should follow somewhere in between the cases of Figures~\ref{fig_MgSi}A and C. 


\subsection{Implications for enstatite chondrites}
Enstatite and ordinary chondrites both show depletion in refractory elements including Mg, but the degree of depletion is higher for enstatite chondrites. Enstatite chondrites have an additional feature, which points to their formation under reduced conditions. This has been suggested from the existence of sulfides and silica \citep{Larimer1979, Grossman2008} based on the assumption that the mineral assemblages of enstatite chondrites represent direct condensates from the nebular gas. Because sulfides do not condense from the initial solar composition, enstatite chondrites must have condensed either from the gas with a higher C/O ratio than the solar ratio \citep{Larimer1979, Grossman2008} or from the gas enriched in interplanetary dust particles \citep{Ebel2005}, both of which produce more reduced condition than the solar composition. The condensates from these compositions, however, also result in the formations of graphite and SiC, neither of which are observed in chondrites. An alternative explanation is that the formation of sulfides is by secondary formation, where oxides reacted with S-rich gas to form sulfides and silica \citep{Lehner2013}. 

Based on our modeling results, the hypothesis of the reaction with S-rich gas may have an astrophysical footing. When planetesimals created from the streaming instability were scattered outwards to the region rich in troilite (FeS) and collided with troilite-rich materials (Figure~\ref{fig_Tdr}D), the reaction between silicates and troilite can proceed and produce the mineralogy observed in enstatite chondrites.
It is noted that although the inward transport of troilite enhances the S/H ratio of the gas phase (Figures~\ref{fig_MgSi}B, \ref{fig_MgSi}D, and \ref{fig_MgSi}F), reaction with the surrounding nebular gas does not produce sulfides (Figure~\ref{fig_Tdr}D). The degree of S-enrichment in the gas phase is only five times the solar abundance and is too low compared to what is necessary for sulfidation reactions \citep{Lehner2013}.

\subsection{Evolution after the onset of streaming instability}
\begin{figure}
\centering
\includegraphics[width=15cm]{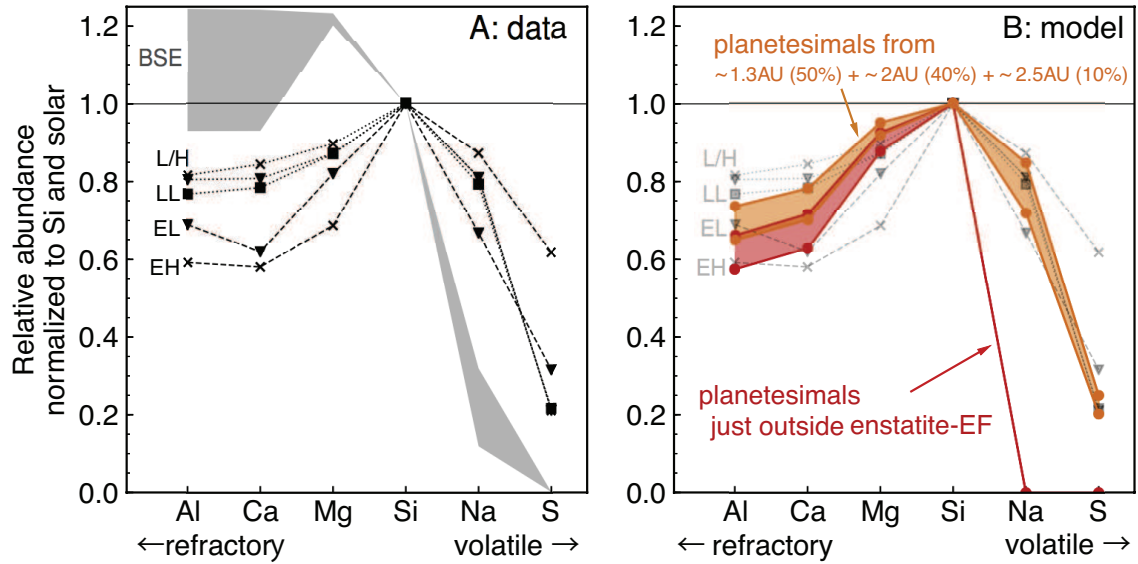}
\caption{Relative abundances of Al, Ca, Mg, Na, and S normalized by Si and by the solar composition for (A) chondrites and BSE, and (B) our modeled results. (A) Data for BSE (gray shaded), ordinary (dotted), and enstatite chondrites (dashed) are shown. Chondrite subgroups are denoted by labels and markers for EH (cross), EL (triangle), H (cross), L (triangle), and LL (square). Data for BSE are taken from \citep{Palme2003, Lyubetskaya2007}, and those for chondrites are from \citep{Wasson1988}. (B) The composition of planetesimals predicted to form outside the evaporation front (EF) of enstatite (red). The initial composition does not contain Na and S because of the high disk temperature, but the volatile content is expected to increase during the later stages by accreting pebbles formed further away from the Sun. The composition of planetary body where 50\% comes from planetesimals formed outside EF enstatite (1.3--1.4~AU) and 40\% from Na-rich pebbles formed at 1.5--2.2~AU, and 10\% from S-rich pebbles formed at 2.3--2.6~AU are shown (orange). The parameters used to create this figure is the same parameters with Figure~\ref{fig_Tdr}.}
\label{fig_comp}
\end{figure}

Planetesimals created outside the evaporation front of enstatite are expected to be depleted in Al, Ca, and Mg, which is consistent with ordinary and enstatite chondrites. Regarding Na, even though this region is enriched in Na, it is too hot for Na-bearing minerals to condense during the onset of the streaming instability. Therefore, the initial composition of planetesimals is likely to be depleted in volatile elements including Na (Figure~\ref{fig_comp}), and planetesimals must have acquired them after their formation. The most likely mechanism of acquisition is pebble accretion \citep{Lambrechts2012}, by which pebbles efficiently accreted onto planetesimals. For planetesimals of 100~km-size typically formed by the streaming instability \citep{Johansen2017}, however, they accrete pebbles only at the later stage of evolution when the surrounding gas dissipates and the disk cools down (Figure~\ref{fig_acc}). Therefore, the composition of planetesimals is unlikely to be changed during the early stage of evolution, and the addition of Na likely happens $\sim$0.1--0.5~Myr after the onset of the streaming instability. By then, volatile-bearing minerals would be stable in the terrestrial region, although some amount of volatiles would have already been dissipated. 

The parent bodies of ordinary and enstatite chondrites may have undergone a similar evolutionary path because the major element compositions of ordinary and enstatite chondrites can be explained by adding volatile-rich pebbles to planetesimals formed outside the evaporation front of enstatite (Figure~\ref{fig_comp}B). In our calculation with $\alpha$=$10^{-3}$, Na- and S-rich pebbles originate from 1.5--3~AU (Figure~\ref{fig_MgSi}A), and they were probably transported inward and accreted to planetesimals during the later stage of evolution. It is noted that such pebbles show weaker Al-, Ca- and Mg-depletions than the initial planetesimals, so their accretion slightly weakens the degree of depletion in refractory elements (Figure~\ref{fig_comp}B). Nevertheless, the final composition can still have larger depletion in refractory elements than Na if the planetesimals are greatly depleted in refractory elements when they initially formed by the streaming instability. 

This highlights that the abundances of refractory and volatile elements are determined by different processes; the depletion of refractory elements, including Al, Ca, and Mg, is due to the preferential loss of refractory minerals before planetesimal formation, whereas that of volatile elements arises from the disk dissipation before the accretion of volatile-bearing pebbles. Through these processes, certain chondrites can have the Na/Al ratio larger than unity, and it is not surprising to have planetesimals that are depleted in refractory elements and are enriched in volatiles at the same time.

\subsection{Evolutionary path to Earth formation}
To explain the depletion of refractory elements observed in ordinary and enstatite chondrites, some mechanism to create nebula-wide fractionation must have operated. Considering that the protoplanetary disk was sufficiently hot to evaporate silicates only in the first million year of evolution \citep{Cassen2001, Chambers2009}, such mechanism has to create the depletion quickly, and parental bodies for those chondrites must also be formed from the depleted region. Our proposal for streaming instability occurring at the evaporation front of enstatite satisfies both of these constraints, and this suggests that planetesimals in the inner region are inevitably depleted in refractory elements. 
On the other hand, carbonaceous chondrites, which show no depletion in refractory elements \citep{Wasson1988}, are likely to be created further away from the Sun. This is consistent with nucleosynthetic isotope variations observed in different chondrite groups \citep{Burkhardt2016}, which likely reflects the heliocentric distance of their source regions.

Enstatite chondrites have likely originated in the terrestrial region, but one of the challenges to forming Earth and enstatite chondrites from the same source \citep{Javoy1995, Javoy2010} is that they have different Mg/Si ratios. Our model, however, suggests that planetesimals with very different Mg/Si ratios can be formed within the same isotopic reservoir. In the region around the evaporation front of enstatite, the Mg/Si ratio changes significantly within a narrow region (Figures~\ref{fig_MgSi}A and \ref{fig_MgSi}C), where an enrichment in Mg (Mg/Si $\sim$1.02--1.05) is observed just inside the region where Mg is depleted. The degree of Mg enrichment is larger for the bulk silicate Earth \citep[][Figure~\ref{fig_comp}A]{McDonough1995, Lyubetskaya2007}, and this suggests that planetesimals formed inside and outside the evaporation front of enstatite, respectively, may have served as the sources for Earth and enstatite chondrites. Also, they are expected to have similar isotopic compositions because the source regions for these two bodies are nearby. Our results thus suggest that similarity in isotopic compositions do not necessarily mean that their bulk compositions also resemble each other. Earth can have an isotopic composition close to enstatite chondrites and at the same time have a bulk composition close to carbonaceous chondrites. One exception is that Earth has a heavier Si isotopic composition than enstatite chondrites. The variation of $\delta^{30}$Si, however, correlates with the Mg/Si ratio and can be explained by the fractionation of forsterite \citep{Dauphas2015}. Therefore, the Mg/Si ratios and the $\delta^{30}$Si values of Earth and enstatite chondrites are both consistent with the preferential loss of forsterite around the evaporation front of enstatite.

The evolutionary path to Earth formation may thus have followed the following steps: First, the radial drift of pebbles in the presence of evaporation fronts caused a depletion in refractory elements around the terrestrial region of the protoplanetary disk. In order to create a large enough depletion seen in ordinary and enstatite chondrites, the inward transport of pebbles should have been sufficiently efficient. Such a high flux of pebbles then resulted in pebble accumulation at the evaporation front of enstatite, naturally leading to the formation of planetesimals by the streaming instability. Whereas planetesimals formed outside the evaporation front remained in their original sizes and became parental bodies for ordinary and enstatite chondrites, those formed just inside the evaporation front may have become the building blocks of Earth. 

\acknowledgments This work was supported in part by the facilities and staff of the Yale University Faculty of Arts and Sciences High Performance Computing Center.


\begin{table}
\centering
\caption{Summary of the representative cases tested in this study. Modeling results are evaluated from three perspectives: a large enough pebble/gas ratio ($>5.0\times 10^{-3}$), the Mg-depletion ($<$0.9), and Al-depletion ($<$0.8), the latter two of which are consistent with ordinary chondrites. All three criteria are met only when turbulence is sufficiently weak ($\alpha\le10^{-3}$), the grain size of pebbles are in the right range (St$_p \sim 0.01$), and the time scale for pebble formation is short enough ($\tau_{d \to p} \le $ 100~yr). See Supporting Information for more cases.}
\begin{tabular}{rrr|ccc}
\large $\alpha$ & \large St$_p$ & \large $\tau_\mathrm{d \to p}$ & \large $\Sigma_p/\Sigma_g$ & \large Mg/Si & \large Al/Si\\
\hline
\large $10^{-2}$ & \large 0.005 & \large $10^2$ & \large $\circ$ & \large $\times$   & \large $\circ$ \\
\large $10^{-2}$ & \large 0.05   & \large $10^2$ & \large $\times$ & \large $\times$ & \large $\circ$ \\
\large $10^{-3}$ & \large 0.0005& \large $10^2$ & \large $\circ$ & \large $\times$   & \large $\circ$ \\
\large $10^{-3}$ & \large 0.005 & \large $10^2$ & \large $\circ$ & \large $\circ$      & \large $\circ$ \\
\large $10^{-3}$ & \large 0.05   & \large $10^2$ & \large $\times$ & \large $\circ$   & \large $\circ$ \\
\large $10^{-3}$ & \large 0.005 & \large $10^3$ & \large $\times$ & \large $\times$ & \large $\circ$ \\
\large $10^{-4}$ & \large 0.005 & \large $10^2$ & \large $\times$ & \large $\circ$ & \large $\circ$ \\
\large $10^{-4}$ & \large 0.05   & \large $10^2$ & \large $\times$ & \large $\circ$ & \large $\circ$ \\
\hline
\end{tabular}
\label{tab_res}
\end{table}

\appendix
\section{Disk evolution model} We constructed a 1+1D disk evolution model, where the temporal and spatial evolution of the chemical compositions of three components, gas, dust, and pebble, are tracked together with the disk temperature, pressure, and their surface densities of the three components. Our model considers mass transport only in the radial direction, but the thermal and compositional structures in the vertical direction are taken into account \citep[][see Supporting Information]{Miyazaki2017}. At each timestep, disk dynamics, dust growth and transport, and Gibbs energy minimization are solved simultaneously, which keeps the thermal and compositional structures of the protoplanetary disk self-consistent. Our model considers a system spanning from 0.05 to 30~AU using 600~bins equally distributed over the entire system, and each bin contains the information on temperature, pressure, and the mass and composition of three components. We use a timestep of 0.01~year to solve for the evolution of surface densities (Equation~(\ref{addiff})), whereas the thermal structure and equilibrium among different phases are updated every 25~years. 

\subsection{Thermal structure} 
Temperature and pressure at the midplane are calculated from energy balance between viscous dissipation and blackbody radiation:
\begin{equation}
	2 \sigma_B T_e^4 = \frac{9}{4} \nu \Sigma \Omega_K^2,
\end{equation}
where $\sigma_B$ is the Stefan-Boltzmann constant and $\Omega_K$ is the local Keplerian angular velocity. The midplane temperature, $T_m$, is calculated from the effective temperature, $T_e$, by integrating the radiative transfer equation from the disk surface to the midplane \citep{Miyazaki2017} (see Supporting Information). The viscosity is scaled using the $\alpha$-prescription \citep{Shakura1973} as $\nu = \alpha c_s^2 / \Omega_K$, where $c_s$ is sound velocity, which is proportional to $\sqrt{T}$. 
The source and degree of turbulence have been an active area of research, and recent findings suggest that a certain degree of turbulence exists throughout the disk. Because the focus of our model is in the early stage of a protoplanetary disk, both magneto-rotational instability and pure hydrodynamic turbulence are considered as the source of turbulent viscosity \citep{Bai2013, Nelson2013, Stoll2014}. We thus adopt $10^{-2}$, $10^{-3}$, and $10^{-4}$ as the possible values of $\alpha$.

\subsection{Dynamic evolution} 
For each species $i$ in the gas and dust phases, its surface density, $\Sigma_i$ is solved using a 1D radial disk evolution model described in Equation~\ref{addiff}. The advection velocity $v_i$ is calculated separately for the gas and dust phases. For the gas phase,
\begin{equation}
	v_i^\mathrm{gas} = - \frac{3}{\sqrt{r} \Sigma} \pd{}{r} \left( \nu \sqrt{r} \Sigma \right)
\end{equation}
is adopted, and for the dust phase, the drift velocity is added to $v_i^\mathrm{gas}$:
\begin{equation} \label{drf}
	v_i^\mathrm{dust} = v_i^\mathrm{gas} + \frac{\mathrm{St}}{1 + \mathrm{St}^2} \frac{1}{\rho_g \Omega_K} \frac{dP}{dr},
\end{equation}
where $\mathrm{St}$ is the normalized stopping time, and $\rho_g$ is the gas density at the disk midplane. The normalized stopping time, also known as the Stokes number, is the ratio of stopping and eddy turnover times, described as a function of grain size, $s$:
\begin{equation}
	\mathrm{St} = \frac{\rho_m}{\rho_g} \frac{s}{c_s} \Omega_K,
\end{equation}
where $\rho_m$ is the material density of dust grains, for which 3300~kg/m$^3$ is adopted in our study. The gas density $\rho_g$ is calculated from the surface density using 
\begin{equation}
	\rho_g = \frac{1}{\sqrt{2 \pi}} \frac{\Sigma_g}{c_s/\Omega_K},
\end{equation}
where $\Sigma_g$ is the gas surface density, calculated as the total surface density of all gas species. The drift velocity of pebbles is calculated using Equation~\ref{drf} as well, but with a different grain size. 

\subsection{Grain size distribution} 
The distribution of grain size of solid particles is necessary to calculate opacity (see Supporting Information) and the advection velocity of the dust phase. Following ref.~\citep{Estrada2016}, we assume that the dust mass distribution can be described by a power law ($f(m) \propto m^{-q}$). Previous studies have shown that nearly constant mass is distributed per decade in a turbulent disk \citep{Weidenschilling2000, Brauer2008, Birnstiel2010}, and we use the exponent of $q=11/6$, which corresponds to slightly more mass being carried by larger grains. The minimum grain size in the distribution is set to 0.1~$\mu$m, whereas the maximum radius, $s_\mathrm{max}$, is characterized by the fragmentation barrier and is described using the following scaling law \citep{Estrada2016}:
\begin{equation}
	s_\mathrm{max} = \frac{Q_*}{\alpha c_s^2},
\end{equation}
where $Q_*$ is the specific energy of fragmentation. A bouncing barrier \citep{Zsom2010} has also been proposed to inhibit the growth, but it is not an impermeable barrier and merely slows the growth of particles. Because particles are likely to grow until they reach the fragmentation barrier given a long enough time, the maximum grain size of $s_\mathrm{max}$ is adopted in our model.
When an evaporation front exists in the vertical direction \citep{Miyazaki2017}, however, frequent evaporation and condensation are likely to occur at the front, inhibiting dust growth. In such a case, we assume that the maximum radius remains at 10~$\mu$m. Dust mass distribution strongly affects dust opacity, which also influences the midplane temperature.

Pebbles, on the other hand, are generated from the dust phase according to the accretionary timescale $\tau_{d \to p}$ \citep{Cassen2001, Ciesla2008}:
\begin{equation}
	\dot{M_p}(r) \propto \frac{M_d}{\tau_{d \to p}},
\end{equation}
where $\tau_{d \to p}$ is assumed to vary inversely proportional to the local Kepler angular velocity ($\tau_{d \to p}(r) = \tau^0_{d \to p} \cdot \Omega_K (1~\mathrm{AU})/\Omega_K (r)$). 
 The timescale can be related to the ratio between dust and pebble masses because the production of pebbles and their loss through radial drift become quickly balanced in most regions. Our treatment of dust growth is simplified, but it still produces a roughly constant pebble/dust ratio in the region not affected by the presence of the evaporation front. This is also seen in models with more detailed grain growth model \citep{Brauer2008, Estrada2016}, and we regard that this implementation is sufficient for understanding the major element chemistry to first order. For $\tau^0_{d \to p}$, we tested values of $10^2$ and $10^3$~yr in our model, corresponding to the pebble/dust mass ratio of $\sim$$10^{-1}$ and $\sim$$10^{-2}$, respectively, for pebbles with St$\sim$0.005. The size of pebbles is estimated from previous studies \citep{Brauer2008, Estrada2016}, and the values of 0.005 and 0.05 are tested in this study.

\subsection{Gibbs energy minimization}
We assume the solar composition for the initial composition, although only major elements (H, C, O, Na, Mg, Al, Si, S, Ca, and Fe) are considered in this study. For gas species, Al, AlOH, Ca, CH$_4$, CO, CO$_2$, Fe, H, H$_2$, H$_2$O, H$_2$S, Mg, Na, NaOH, O$_2$, Si, SiO, and SiS are considered, and for mineral species, corundum (Al$_2$O$_3$), melilite (solid solution between gehlenite (Ca$_2$Al$_2$SiO$_7$) and Ca$_2$MgSi$_2$O$_6$ (akermanite)), 
olivine (solid solution between forsterite (Mg$_2$SiO$_4$) and fayalite (Fe$_2$SiO$_4$)), orthopyroxene (solid solution between enstatite (MgSiO$_3$) and ferrosilite (FeSiO$_3$)), spinel (MgAl$_2$O$_4$), anorthite (CaAl$_2$Si$_2$O$_8$), diopside (CaMgSi$_2$O$_6$), albite (NaAlSi$_3$O$_8$), metallic iron (Fe), troilite (FeS), and graphite (C) are considered. Thermodynamical data for silicate phases are taken from Robie \& Hemingway (1995), and the rest of the phases are from the JANAF Thermochemical Tables. We use the method of ref.~\citep{Miyazaki2017} to perform Gibbs energy minimization. It is noted that orthopyroxene and olivine in Figures~\ref{fig_Tdr}C and \ref{fig_Tdr}D are both dominated by their Mg-end members. The solid/gas ratio in our model is not high enough to stabilize significant amount of fayalite and ferrosilite in the terrestrial region.






\newpage

\section{Supporting Information}

\subsection{Vertical temperature structure}
Thermal and compositional structures are kept self-consistent by solving the vertical structure of the protoplanetary disk as described in \citep{Miyazaki2017}. Temperature at the disk surface is calculated by balancing viscous dissipation and blackbody radiation (see Methods), and the temperature of the disk interior is calculated from the surface towards the midplane by solving the radiative heat transfer equation:
\begin{equation} \label{dTdz}
	\frac{dT}{dz} = - \frac{3 \overline{\kappa_R} \rho_g}{16 \sigma_B T^3} F_z
\end{equation}
where $z$ is the height from the midplane, $\overline{\kappa_R}$ is the Rosseland mean opacity, and $F_z$ is the vertical radiative flux. Temperature variation in the vertical direction affects the gas and dust compositions at different heights, resulting in a change in the Rosseland mean opacity. Therefore, the mass and composition of dust grains are solved at each height simultaneously with temperature using Gibbs energy minimization, and the calculated dust property is used to determine opacity. The radiative flux $F_z$ is also calculated downwards from the surface to the midplane together with Equation~(\ref{dTdz}) using the following equation:
\begin{equation}
 	\frac{dF_z}{dz} = \frac{9}{4} \nu \rho_g \Omega_k^2,
\end{equation}
with an boundary condition of $F_z = 0$ at the midplane ($z=0$). We assume that the maximum grain size in regions where minerals evaporate is reset to the initial nucleation size of 1~$\mu$m. This allows us to include the effect of evaporation front on the vertical thermal structure without modeling mass transport in the vertical direction \citep{Miyazaki2017}. 

The Rosseland mean opacity of dust particles is calculated as a function of the dust/gas ratio, grain size, and temperature, and we use the opacity data obtained in \citep{Miyazaki2017}. As a first-order estimate, the opacity of silicate is used for all dust compositions in this study. When grains are larger than 1~$\mu$m, their opacity is roughly inversely proportional to the grain size, and the effect of pebbles on opacity is negligible compared to that of dust with small grain sizes.

 
\subsection{The implementation of porosity}
We tested different porosities for pebbles, where a higher porosity results in a larger surface area and thus enhances re-condensation onto pebbles. When pebbles are highly porous, the streaming instability may be possible even with a longer accretionary timescale or a larger pebble size. Porosity is described using the exponent $q$ in the following equation in our model:
\begin{equation}
	\frac{M_p}{M_d} = \frac{S_p}{S_d} \left( \frac{r_p}{r_d} \right)^q,
\end{equation}
where $M$ denotes the total mass, $S$ is the total surface area, and $r$ is the grain radius of each phase. Subscripts $p$ and $d$ indicate the pebble and dust phases, respectively. Spherical pebbles correspond to $q=1$, whereas porous pebbles could have the value as low as 0. We tested 1, 1/2, and 1/4 for the possible values of $q$, and the pebble/gas ratio increases around the evaporation front of enstatite as a smaller value of $q$ is adopted. In regions away from the evaporation front, however, the effect of re-condensation is insignificant, and the pebble/gas ratio does not change much with $q$.

\subsection{Planetesimal formation by gravitational instability} In order to overcome various growth barriers, we assume that planetesimals are created directly from pebbles through gravitational instabilities. The streaming instability \citep{Youdin2005, Johansen2007, Johansen2014} is considered in this paper, which concentrates pebbles into dense filaments, which can further collapse into planetesimals under self-gravity. The other potential mechanism, turbulent concentration \citep{Cuzzi2001, Cuzzi2008}, is not considered here because the degree of turbulent concentration has been recently suspected to be lower than initially proposed \citep{Pan2011}. 

Turbulent concentration has been considered as a possible mechanism to create planetesimals especially in the inner disk because, without the re-condensation of evaporated dust grains onto pebbles, the streaming instability has been considered unlikely. The streaming instability requires a large pebble/gas ratio, but the amount of solid in the inner region is smaller than in the outer region because of the lack of icy particles. This issue may be circumvented by turbulent concentration because turbulence may locally concentrate pebble-size grains by a factor of 100 or more, resulting in gravitational instability \citep{Cuzzi2001, Cuzzi2008}. Because this mechanism does not require an enhancement of the dust/gas ratio, turbulent concentration could potentially create planetesimals at anytime and any place of the protoplanetary disk.

\begin{figure}
\centering
\includegraphics[width=17cm]{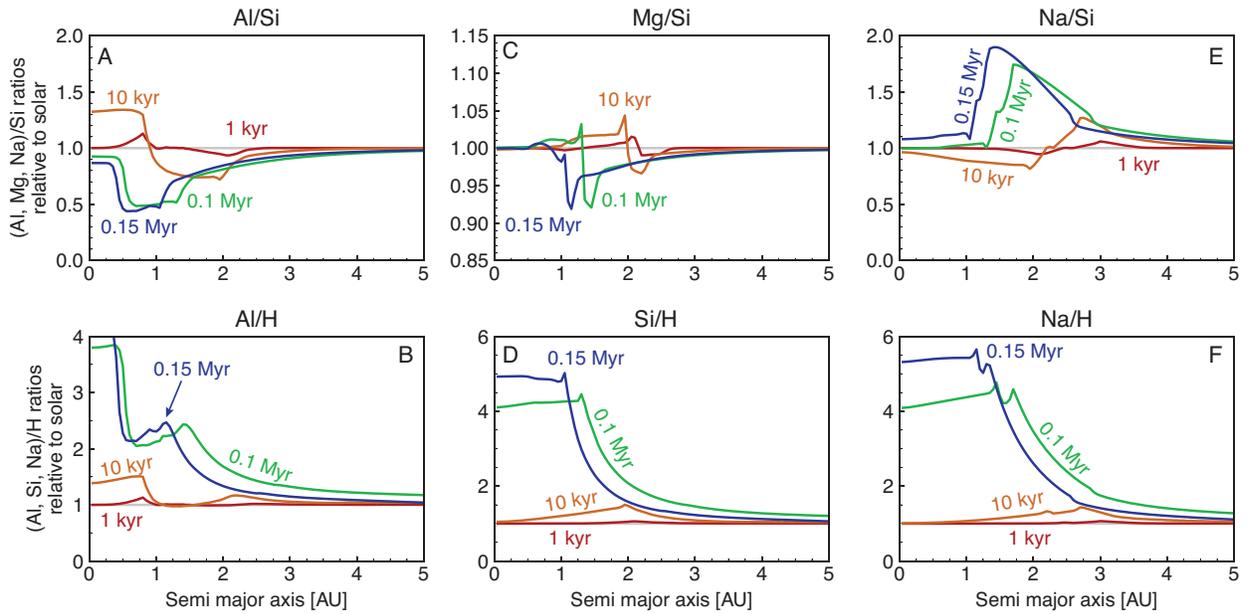}
\caption{The evolution of the relative abundance profiles of (A) Al/Si, (C) Mg/Si, and (E) Na/Si normalized by the solar composition, and (B) Al/H, (D) Si/H, and (F) Na/H normalized by the solar composition. Colors denote 1~kyr (red), 10~kyr (orange), 0.1~Myr (green), and 0.15~Myr (blue). The parameters used to create this figure are the same with Figures~1, 2A, 2B, and 3B in the main text. Gray lines in all panels denote unity, corresponding to an unfractionated value.
}
\label{fig_MgSi_evo}
\end{figure}

In our model, the region where refractory elements are depleted, especially where Mg is depleted, exists only in a narrow region, and such a region moves towards the inner disk as the disk evolves (Figure~\ref{fig_MgSi_evo}). If gravitational instability induced by turbulent concentration were the mechanism to create parental bodies for ordinary and enstatite chondrites, it also implies that the source regions for those chondrites are likely to be distributed throughout the disk. In such a case, the source regions for ordinary and carbonaceous chondrites have a high probability of overlapping. This, however, would not explain the nucleosynthetic isotope variations observed in different chondrite groups~\citep{Burkhardt2016}. We thus suggest that from the perspective of major element chemistry, the streaming instability is a more likely mechanism to explain the nucleosynthetic anomalies.

\begin{figure}
\centering
\includegraphics[width=13cm]{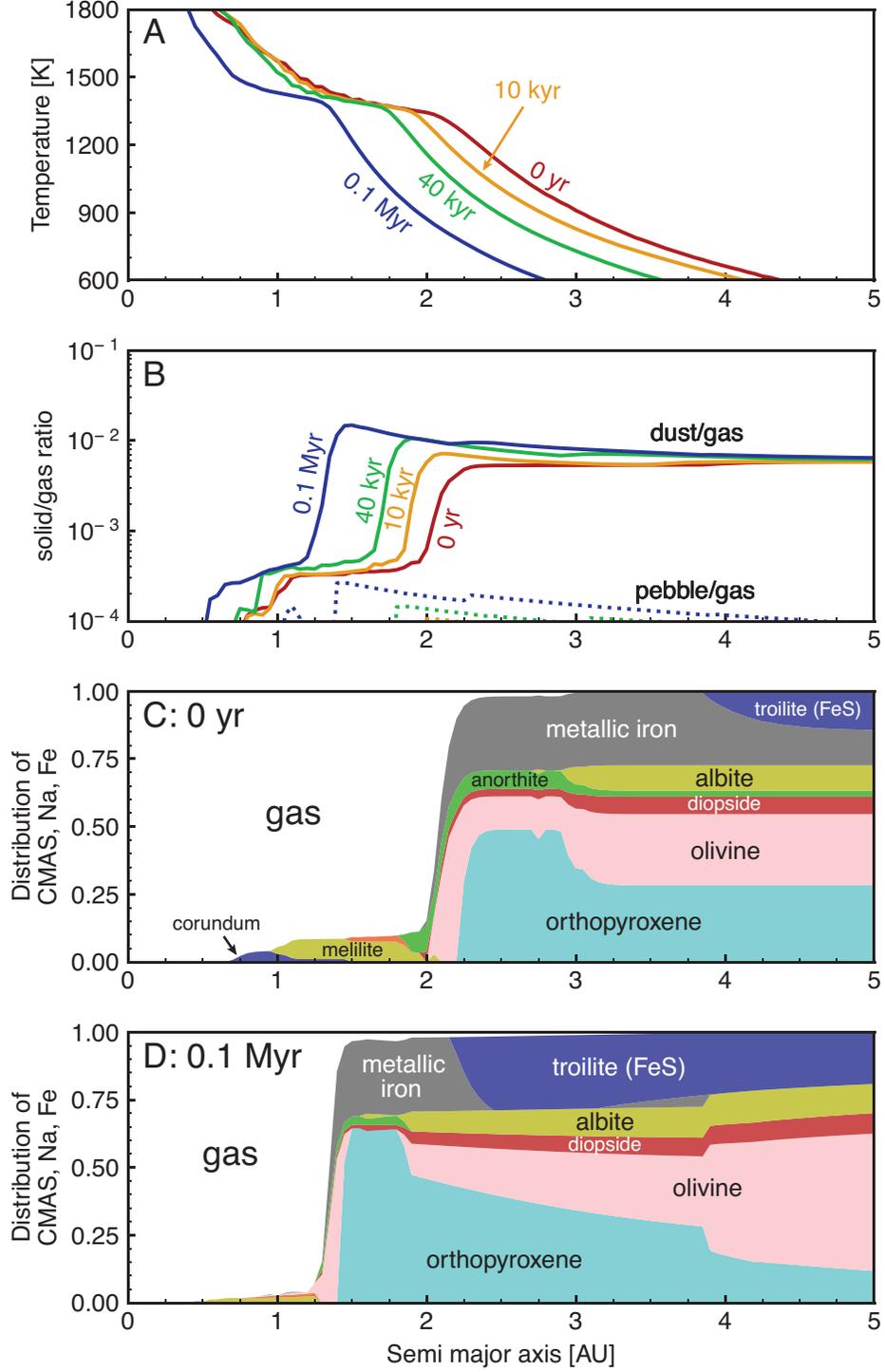}
\caption{The same as Figure~1 but for $\alpha$=$10^{-3}$, $\mathrm{St}_p$=0.05, and $\tau_{d \to p}$=100~yr, which are the same parameters used in Figures~2C and 2D. (A, B) Colors denote 0 kyr (red), 3~kyr (orange), 10~kyr (green), and 40~kyr (blue). Solid and dotted lines in (B) indicate the ratios of dust/gas and pebble/gas, respectively. (C, D) Predicted mineral assemblages of the system at (C) $t$=0 yr and (D) $t$=40~kyr. The relative molar amount of dust-composing elements, Na, Mg, Al, Si, Ca, and Fe, in the solid phase is plotted as a function of heliocentric distance, where colors denote different mineral phases: corundum (Al2O3, dark blue), melilite (Ca2Al2SiO7, yellow), orthopyroxene (opx, (Mg,Fe)SiO3, cyan), olivine ((Mg,Fe)2SiO4, pink), diopside (CaMgSi2O6, red), anorthite (CaAl2Si2O8, green), albite (NaAlSI3O8, yellow), metallic iron (Fe, gray), and troilite (FeS, purple).}
\label{fig_Tdr_32294}
\end{figure}

\begin{figure}
\centering
\includegraphics[width=11cm]{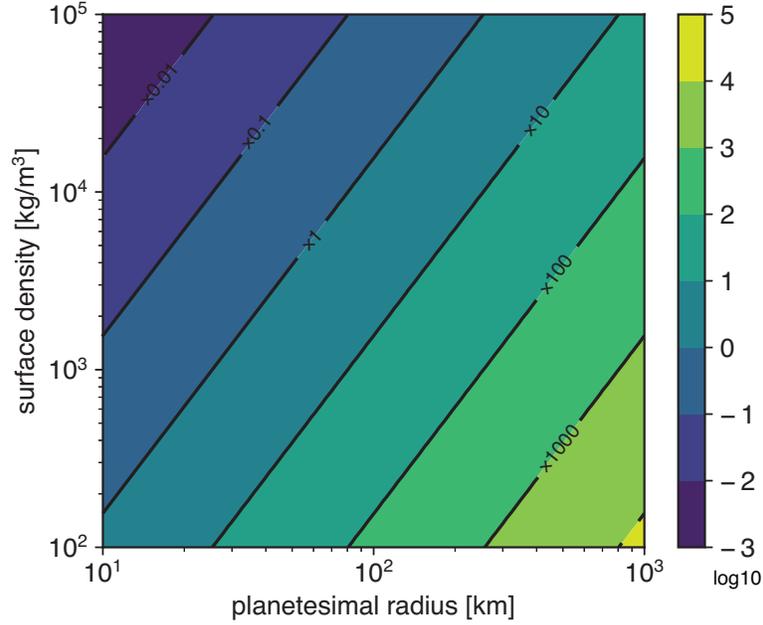}
\caption{The ratio between the Bondi accretion and planetesimal radii as a function of the surface density and planetesimal radius. When a Bondi accretion radius is 100 times the planetesimal radius, for example, the radius of accretion cross section is increased by 100 times from its original planetesimal size. Planetesimals can collect pebbles from a larger region as their size becomes larger and as the surface density of the surrounding gas decreases. The Bondi accretion radius is described as $GM_p / \Delta v$, where $G$ is the gravitational constant, $M_p$ is planetesimal mass, and $\Delta v$ is difference in the orbital velocity of the gas and dust phases. The parameter used to create this figure is the same with Figure~1 in the main text.
}
\label{fig_acc}
\end{figure}

\end{document}